\documentclass[aps,prl,superscriptaddress,twocolumn,letterpaper,amssymb,amsmath,floatfix]{revtex4-1}

\pdfoutput=1

\usepackage{graphicx}
\usepackage{youngtab}

\begin{document}

\title{Design local spin models for Gutzwiller-projected parton wave functions}
\author{Jia-Wei Mei}
\affiliation{Perimeter Institute for Theoretical Physics, Waterloo, Ontario, N2L 2Y5 Canada}
\author{Xiao-Gang Wen}
\affiliation{Perimeter Institute for Theoretical Physics, Waterloo, Ontario, N2L 2Y5 Canada}
\affiliation{Department of Physics, Massachusetts Institute of Technology, Cambridge, Massachusetts 02139, USA}
\date{\today}
\begin{abstract}
We introduce a method to design a local spin Hamiltonian to realize a
Gutzwiller-projected parton wave functions (GPWF) as its ground state. For
example, the Dirac spin liquid (DSL) state is quite close to the true ground
state of the spin-1/2 Heisenberg model on kagome lattice.  We examine what kind
of perturbations we should add in order to drive the DSL to more stable  chiral
spin liquid (CSL), valence bond solid (VBS) or Gutzwiller-projected spin Hall
(GSH) states.  We compute the two-body reduced-density-matrices (2-RDM) of
GPWFs for those target states, and compare them to the 2-RDM of the DSL. This
allows us to design local spin models with only two-body interactions that may
realize those interesting target states.  Our results agree very well with
recent numerical calculations for CSL on kagome lattice.  We also study spin-1
systems on kagome lattice, and design local spin models that may realize CSL,
VBS and symmetry-protected topological (SPT) states.  Our work establishes a
directional guide for further numerical simulations.

\end{abstract}
\maketitle

Quantum entanglement has been becoming an important concept to understand and
classify the quantum many-body system. It precisely rephrases ``topological
order'' as long-range entanglement.\cite{Levin2006,Kitaev2006a} The topological
ordered state is not smoothly connected to a direct product state by any local
unitary transformation.\cite{Chen2010} It has new topological quantum numbers,
such as non-trivial ground state structures  and fractional
excitations.\cite{Wen1989,Wilczek1984,Wen1990,Wen1990a,Laughlin1983,Wilczek1984,Arovas1984}
In the presence of symmetry, symmetry-protected topological (SPT) state,
although short-range entangled, cannot be locally deformed into a direct
product state via local unitary transformations that preserve the
symmetry.\cite{Chen2012,Qi2012,Chen2013} Several exactly solvable lattice
models are constructed to realize topological orders\cite{Kitaev2003,Wen2003,Levin2005,Kitaev2006} and
SPT states\cite{Chen2011,Walker2011}. Kitaev $XYZ$ model on the honeycomb
lattice\cite{Kitaev2003} is an elegant example to use spin Hamiltonian with
only two-body interactions to realize a particular topologically ordered state.
However, it is a challenging task to find an exactly solvable local spin model
that has a generic topologically ordered or a SPT ordered ground state.  For a
two-dimensional (2D) spin model that is not exactly soluble, expensive
numerical methods, e.g., exact diagonalization (ED)\cite{Luescher2009},
projected entangled pair states (PEPS) or tensor-network\cite{Verstraete2008}
and density matrix renormalization group (DMRG)\cite{Stoudenmire2012}, are
required to access the ground state. 

On the other hand, Gutzwiller projective parton construction is a powerful
theoretical approach, that allows us to construct the wave functions of many
interesting and highly non-trivial topological states for strongly correlated
bosonic or spin systems.\cite{Wen1991,Wen1999,Ye2013,Lu2012} In this letter, we
will make a good use of those rich results.  We will carry out a reverse
engineering: to design a local spin Hamiltonian that realize those topological
states described by Gutzwiller-projected parton wave functions (GPWF).

Kitaev $XYZ$ spin model\cite{Kitaev2003} and Wen plaquette model\cite{Wen2003}
are special examples, in which the exact ground states are given by GPWFs.  For
a general GPWF, we might be able to find a very complicated Hamiltonian so that
the GPWF is the exact ground state.  However, real spin systems in general only
have strong two-body interactions.  In this paper, we will try to design a
model Hamiltonian with only two-body interactions, which may optimally realize
a interesting GPWF as the ground state.

To design the two-body Hamiltonian, we investigate two-body
reduced-density-matrix (2-RDM) of GPWFs. We trace out all configurations
except states on two sites of local links, e.g., nearest neighbor (NN), second
NN (2NN) and third NN (3NN) bonds. If the 2-RDMs contain sectors with small
weights (small eigenvalues), it implies that the GPWF is not frustrated for
two-body interactions, and we can design a two-body Hamiltonian to project out
those low-weight sectors. The designed Hamiltonian may realize the  GPWF as its
ground state.  The similar idea can be found in the AKLT model
construction\cite{Affleck1987}, where the spin-2 sector in 2-RDM has a
vanishing weight and is projected out by the designed Hamiltonian.

In this letter, we will take a slightly different approach. We will start with
a Dirac spin liquid (DSL) state which is known to be close to the ground state
of a simple spin model.  We then try to design a perturbation to the original
model that may optimally drive the DSL to a more stable state in its
neighbor where the Dirac points are gapped.
%
%

The $\pi$-flux state is an extensively studied DSL for the spin-1/2 system on the 2D
square lattice.\cite{Affleck1988}  It is the parent state of many different
kinds of states (e.g., antiferromagnetic, $d$-wave superconducting and
pseudogap states) in high $T_c$ cuprates.\cite{Lee2006} DSL is proposed to be
the ground state of spin-1/2 Heisenberg model on the Kagome
lattice.\cite{Ran2007} Different DSL states are also constructed on the
honeycomb lattice.\cite{Clark2011,Albuquerque2011,Corboz2012} 

To design a perturbation that drives the DSL to one of its neighbors, we compute
the 2-RDMs on several bonds for both  DSL and its neighbor.  If the two
2-RDMs are about the same on a bond, then the two-body interaction on that bond
will not be helpful to drive the DSL to its neighbor.  On the other hand, if
the two 2-RDMs are different for some bonds, then, we can design a two-body
perturbation on that bond to favor the neighboring state.  In particular, if
the 2-RDM for the neighboring state has more weight in the high weight sector
and less weight in the low weight sector, it will imply that the neighboring
state is less frustrated for the designed two-body Hamiltonian, and is likely
to be realized.


We apply such an approach to DSL states on kagome lattice for
spin-1/2 SU(2) and spin-1 SU(3) systems. The DSL states are very close to the
ground states for SU(N) ($N=2,3$) spin Hamiltonians with only $J_1$ term on NN
bonds in Eq.(\ref{eq:sun}). We want to drive the DSL to the chiral spin
liquid (CSL), valence bond solid (VBS) and Gutzwiller-projected spin Hall (GSH)
for spin-1/2 systems, and to CSL, VBS and SPT states for spin-1 systems, respectively. From
the variations of 2-RDM, we design the possible local spin models for those
states.

The SU(N) spin operator $S_\beta^\alpha$ has the parton (Schwinger-fermion)
representation, $S_\beta^\alpha(i) = f_{i\alpha}^\dag f_{i\beta}$. Here
$f_{i\alpha}$ is the fermionic parton operator on $i$-site on the lattice and
$\alpha = 1, 2,\cdots,N$ is the spin (flavor) index. On every site we have the
single-occupation constraint, $\sum_\alpha f_\alpha^\dag(i) f_\alpha(i) = 1$.
The GPWF is written as $|\Psi_G\rangle=\sum_{\{R\}}\psi(\{R\})|\{R\}\rangle$,
where the determinate $\psi(\{R\})$ is the wave amplitude for spin (flavor)
configuration basis $|\{R\}\rangle$ on the lattice in the ground state of the
mean field Hamiltonian,  $H_{\text{MF}}=\sum_{\langle ij\rangle}(t_{ij}^\alpha f_{i\alpha}^\dag f_{j\alpha} + u_{ij}^{\alpha\beta}f_{i
  \alpha}^\dag f_{j\beta}^\dag) + \text{H.C.}$. By definition, 2-RDM on sites $i$ and $j$ is written as
\begin{eqnarray}
  \rho(ij)&&=\text{tr}_{\{R\}}'|\Psi_G\rangle\langle\Psi_G|=\sum_{\{ij\},\{ij\}'}|\{ij\}\rangle\langle\{ij\}'|\nonumber\\
\sum_{\{R\}}&&\frac{\psi^*(\{ij\}'\otimes\{R\}/\{ij\}^R)}{\psi^*(\{R\})}\delta_{\{ij\},\{ij\}^R}|\psi(\{R\})|^2,
\end{eqnarray}
where $\{ij\}$ is the spin configuration on sites $i$ and $j$. $\text{tr}'$ is
the trace running over all spin configurations except two states on sites $i$
and $j$.  $\{ij\}'\otimes\{R\}/\{ij\}^R$ is the configuration by replacing
$\{ij\}^R$ in $\{R\}$ by $\{ij\}'$. 2-RDM can be simulated by using the
standard Monte Carlo method according to the weight $|\psi(\{R\})|^2$.

\begin{figure}[t]
  \begin{center}
    \includegraphics[width=\columnwidth]{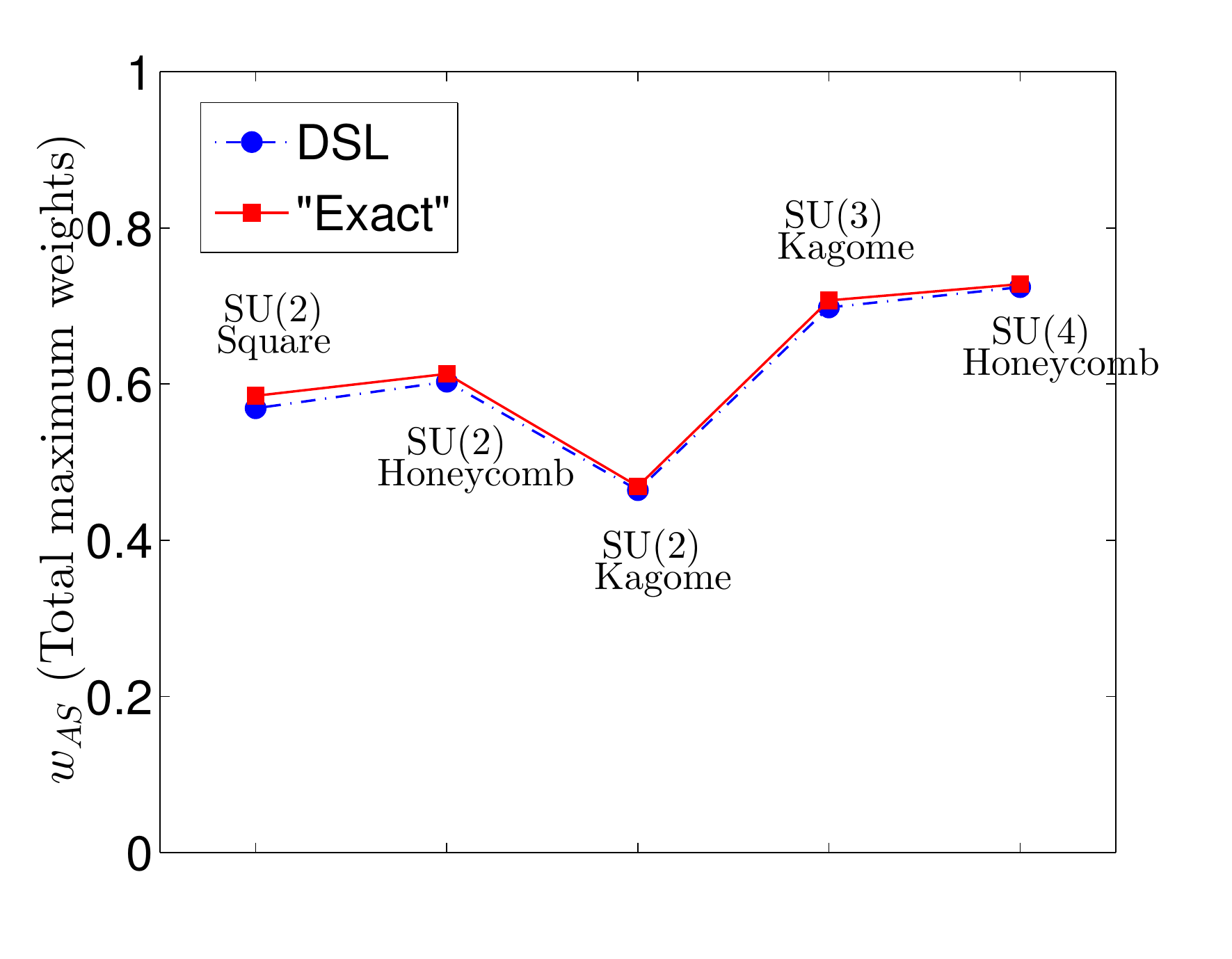}\\
\includegraphics[width=0.65\columnwidth]{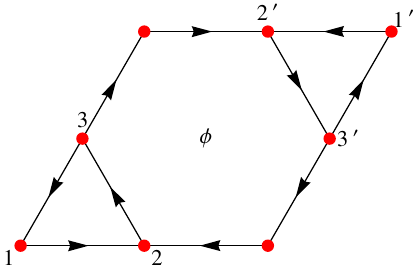}
  \end{center}
  \caption{(a). The anti-symmetric sector (AS) of 2-RDM has $\frac{N(N-1)}{2}$-fold degenerate maximum weights. The total maximum weights $w_{AS}$ on \textit{NN bonds} are shown here.
Blue solid circles are Monte Carlo results for different SU(N) DSL states on different lattices; red solid squares are extracted from  the numerical ``exact'' ground state energies of $J_1$ SU(N) models,  e.g., SU(2) on the square lattice\cite{Trivedi1989}, honeycomb lattice\cite{Albuquerque2011}, kagome lattice\cite{Waldtmann1998} and SU(3) on the kagome lattice\cite{Corboz2012a} and SU(4) on the honeycomb lattice.\cite{Corboz2012} (b). The kagome lattice. The NN bonds (e.g., $\langle12\rangle$) are defined as the sides of triangles $\triangle_{123}$ and $\triangledown_{1'2'3'}$. 2NN (e.g., $\langle\langle 23'\rangle\rangle$) and 3NN (e.g., $\langle\langle\langle 22'\rangle\rangle\rangle$) bonds are in the hexagon. }
  \label{fig:dsl}
\end{figure}

DSL has $u_{ij}=0$ in the mean field Hamiltonian. The complex phase of $t_{ij}$
brings a flux pattern for the parton hopping on the lattice. SU(2) DSL has 
$\pi$ flux in the square plaquette on the square lattice and in the hexagon on
the kagome lattice, respectively.\cite{Affleck1988,Ran2007} There is also $\pi$
flux in the hexagon for SU(4) DSL on the honeycomb lattice.\cite{Corboz2012}
Flux is zero for SU(2) and SU(3) DSL states on the honeycomb
lattice\cite{Albuquerque2011,Clark2011} and kagome lattice, respectively. The
2-RDM of DSL states has the SU(N) anti-symmetric representation {\tiny{
\yng(1,1)}} and symmetric one {\tiny{\yng(2)}} with the dimensions
$\frac{N(N-1)}{2}$ and $\frac{N(N+1)}{2}$, respectively. We use the notations,
$w_{\text{AS}}=\sum_{i=1}^{N(N-1)/2}w_{\text{AS}}^i$ and
$w_{\text{S}}=\sum_{i=1}^{N(N+1)/2}w_{\text{S}}^i$, for total anti-symmetric
and symmetric weights, respectively. The normalization is
$w_{\text{AS}}+w_{\text{S}} = 1$. $w_\text{\text{AS}}$ is dominant in 2-RDM on
NN bonds for these DSL states as shown in Fig. \ref{fig:dsl} (a).

The SU(N) spin model on the 2D lattice is written as
\begin{eqnarray}
  \label{eq:sun}
  H_0 = J_1 \sum_{\langle ij\rangle}P_{ij}+J_2\sum_{\langle\langle ij\rangle\rangle}P_{ij}+J_3\sum_{\langle\langle\langle ij\rangle\rangle\rangle}P_{ij}+\cdots,
\end{eqnarray}
where $J_1$, $J_2$ and $J_3$ are exchange couplings on NN, 2NN and 3NN bonds. $P_{ij}$ is the SU(N) permutation operator, $P_{ij}=S_\beta^\alpha(i)S_\alpha^\beta(j)$,
which swaps two quantum states on bonds. In Fig. \ref{fig:dsl} (a), we collect
some numerical results of $w_{AS}$ of 2-RDM on NN bonds for  ground states of
SU(N) $J_1$ spin model, where $w_{AS}=(1-E_b^1/J_1)/2$ with $E_b^1$ the ground
state energy. DSL states are close to ground states of  SU(N) $J_1$ spin models as seen in Fig. \ref{fig:dsl} (a).



The mean field Hamiltonian of SU(2) DSL on the kagome lattice is written as
\begin{eqnarray}
  H_{\text{DSL}}=-t\sum_{\langle ij\rangle\alpha}e^{i\varphi_{ij}}f_{i\alpha}^\dag f_{j\alpha}+\text{h.c.},
\end{eqnarray}
where $\varphi_{ij}$ brings $\pi$ flux in the hexagon and zero flux in
triangles $\triangle_{123}$ and $\triangledown_{1'2'3'}$.  The CSL state is
obtained by adding extra phase $\theta$ on the directed links on kagome lattice
(Fig. \ref{fig:dsl} (b)). The fluxes  are now $3\theta$ in triangles
$\triangle_{123}$ and $\triangledown_{1'2'3'}$ and $\pi-6\theta$ in the
hexagon. With $\theta=0$ fixed, the VBS state is obtained by varying the ratio
of hopping amplitudes $t_2/t_1$, where $t_1$ and $t_2$ are for bonds on
triangles $\triangle_{123}$ and $\triangledown_{1'2'3'}$, respectively. 
\begin{figure}[b]
  \begin{center}
    \includegraphics[width=0.95\columnwidth]{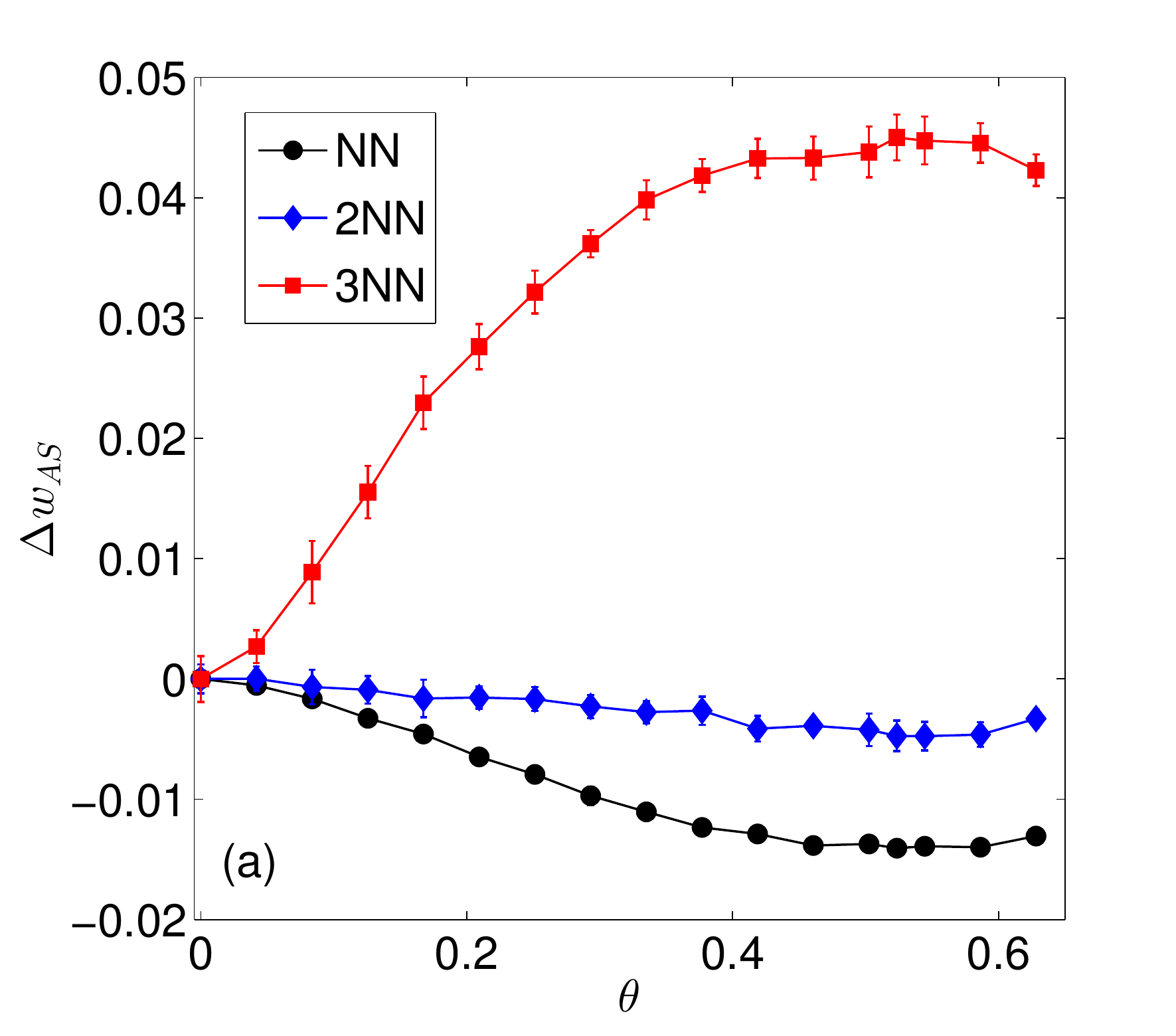}\\
    \includegraphics[width=\columnwidth]{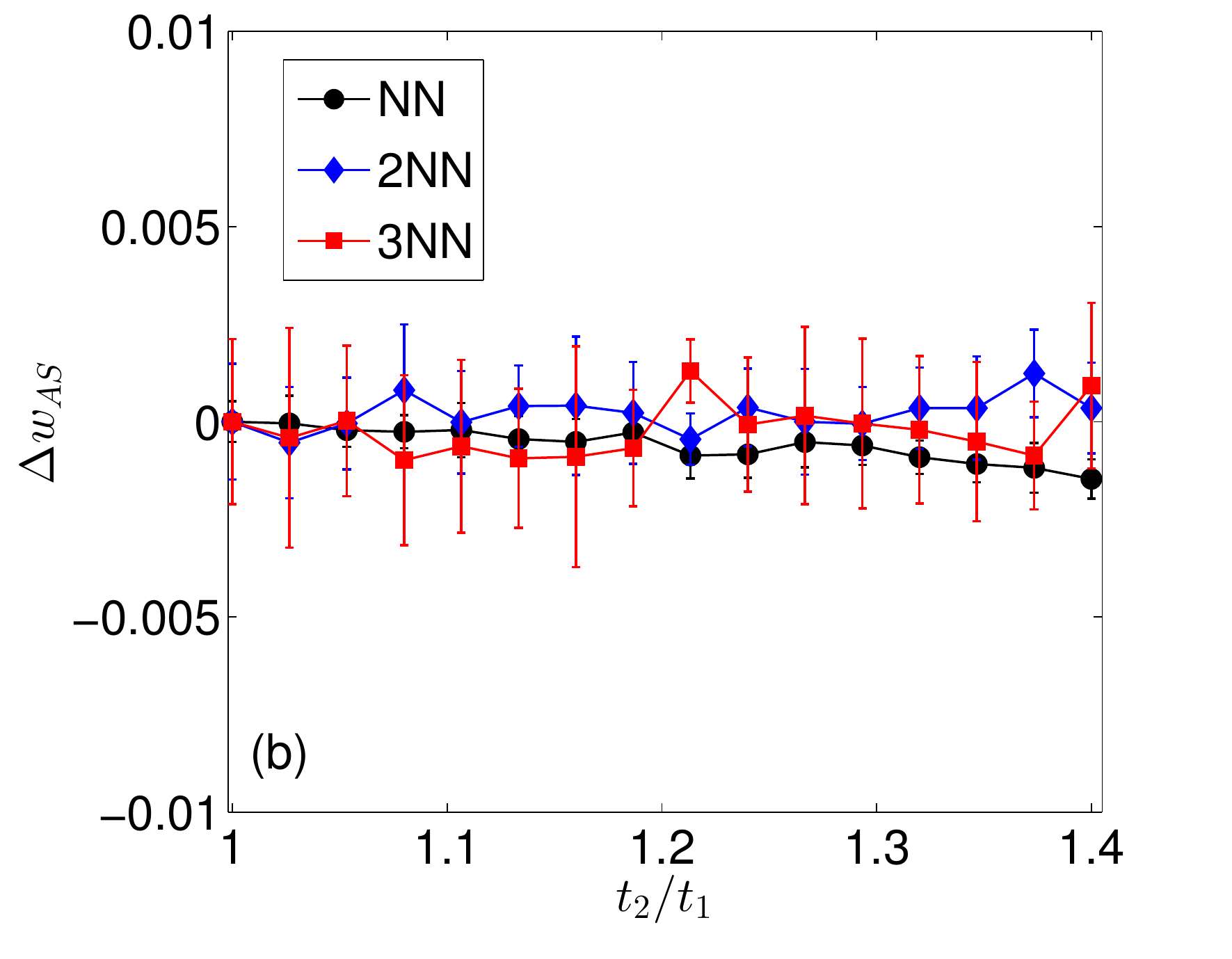}
  \end{center}
  \caption{$\theta$ and $t_2/t_1$-dependent variations of total maximum
anti-symmetric weights $\Delta w_{AS}=w_{AS}-w_{AS}^0$ for SU(2) CSL  (a) and
VBS (b), respectively, on NN, 2NN and 3NN bonds. $\Delta w_{AS}>0$ for SU(2)
CSL increases significantly only on 3NN bonds with varying the wave function
parameter $\theta$. So only $J_3$ spin term on 3NN bonds can favor
CSL, which agrees very well with Refs. \onlinecite{Gong2013, He2014}.}
  \label{fig:su2}
\end{figure}

The anti-symmetric (singlet) weight of 2-RDM for the SU(2) DSL is
$w_{\text{AS}}^0(\text{NN})=0.4645(6)$, $w_{\text{AS}}^0(\text{2NN})=0.261(1)$,
$w_{\text{AS}}^0(\text{3NN})=0.233(2)$.  Both CSL and VBS states have the SU(2)
spin rotation symmetry. With increasing wave function parameters ($\theta$ or
$t_2/t_1$), the responses of 2-RDM on different bonds for different states,
$\Delta w_{AS}=w_{AS}-w_{AS}^0$, behave differently as shown in Fig.
\ref{fig:su2}.  

$w_{\text{AS}}$ of CSL on 3NN bonds is very sensitive to the variation of
$\theta$ and reaches the maximum at $\theta_c=\pi/6$ where the gap in the mean
field Hamiltonian is largest. Meanwhile, VBS is not sensitive to the variation
of $t_2/t_1$. Since $w_{\text{AS}}$ of CSL increases significantly on 3NN bonds
as varying $\theta$, we can add $J_3$ term to the spin model to stabilized the
CSL state. The spin-1/2 SU(2) permutation operator is the Heisenberg term, $ P_{ij}^{S=1/2}=2(\mathbf{S}_i\cdot\mathbf{S}_j+\frac{1}{4})$.
CSL is the potential ground state for $J_1$-$J_3$ Heisenberg model on the
kagome lattice. The estimated critical $J_3$ is around $J_3/J_1 > 0.3$. The CSL
state was already found in DMRG results on $J_1$-$J_2$-$J_3$ Heisenberg model
on the kagome lattice in Refs. \onlinecite{Gong2013,He2014}. $J_2$ term on 2NN
bonds is also included in Refs. \onlinecite{Gong2013,He2014}. From the variation
of 2-RDM, $J_2$ term is not important (even not favored) for the CSL state.

\begin{figure}[b]
  \begin{center}
\includegraphics[width=\columnwidth]{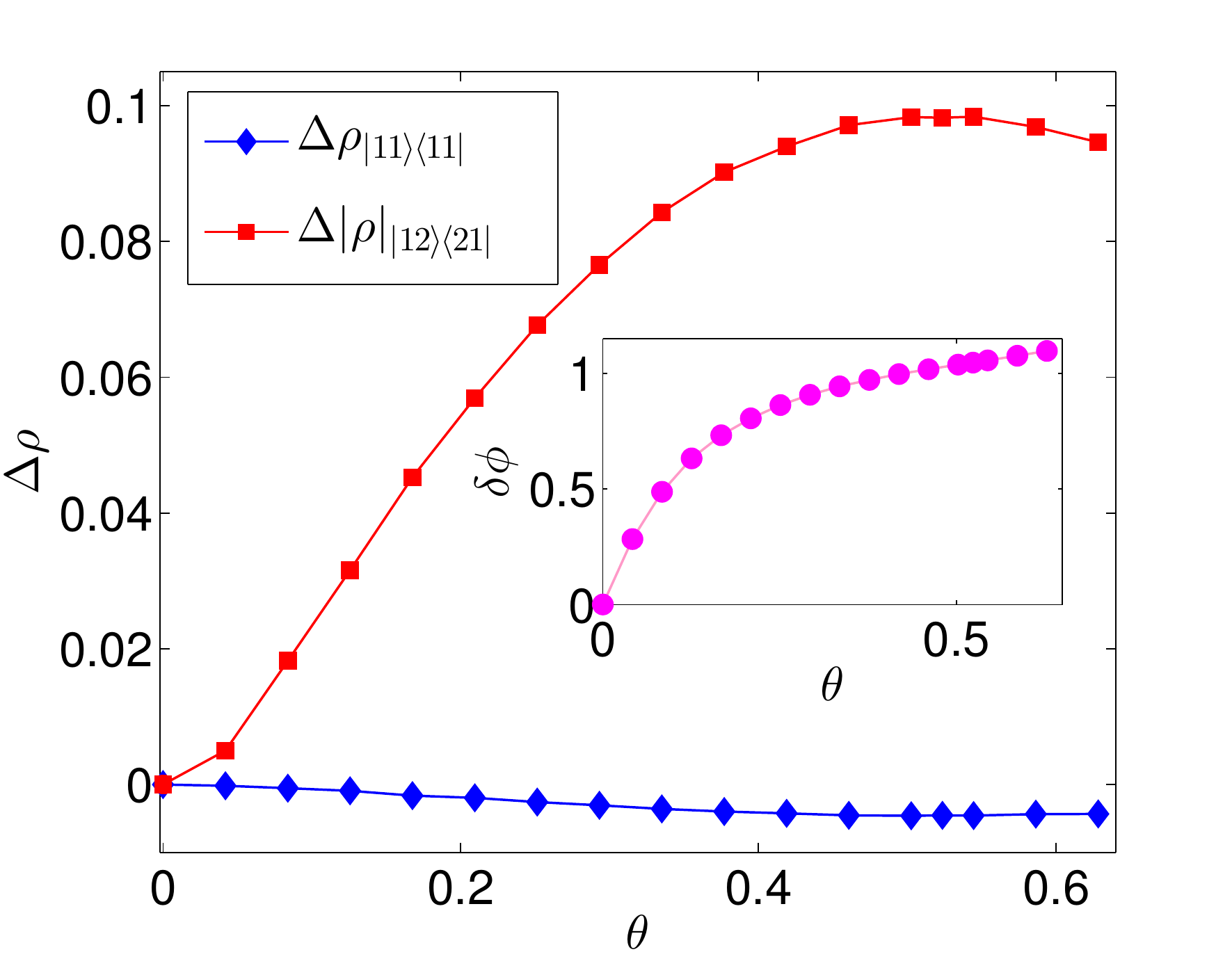}
  \end{center}
  \caption{The $\theta$-dependence of variations of 2-RDM for the GSH state on NN bonds is defined in Eq. (\ref{eq:gshrdm}). The amplitude $|\rho|_{|12\rangle\langle21|}$ and phase $\delta\phi$ increase significantly implying that GSH is stabilized by DM interactions.\cite{Cepas2008} }
  \label{fig:gsh}
\end{figure}

A planar anti-ferromagnetic state can be modeled as the GSH state with the mean
field Hamiltonian\cite{Hermele2008}
\begin{eqnarray}
  H=-t\sum_{\langle ij\rangle}e^{i\phi_{ij}+i\theta_\alpha}f_{i\alpha}^\dag f_{j\alpha}+\text{h.c.},
\end{eqnarray}
where different flavors see opposite flux, $\theta_1=-\theta_2=\theta$. 2-RDM of GSH has two independent components of 2-RDM, $\rho_{|11\rangle\langle11|}=\rho_{|22\rangle\langle22|}$ and $\rho_{|12\rangle\langle21|}$. The notations for variations of 2-RDM are as follows
\begin{eqnarray}
\label{eq:gshrdm}
 \Delta\rho_{|11\rangle\langle11|} &=&-( \rho_{|11\rangle\langle11|} - \rho^{0}_{|11\rangle\langle11|}),\nonumber\\
\Delta|\rho|_{|12\rangle\langle21|} &=& |\rho|_{|12\rangle\langle21|} - |\rho|^{0}_{|12\rangle\langle21|},\nonumber\\
\delta\phi&=&\arg(-\rho_{|12\rangle\langle21|}). 
\end{eqnarray}
$\rho^{0}_{|11\rangle\langle11|} = 0.1786(2)$, $\rho^{0}_{|12\rangle\langle21|}=-0.143(1)$ are the components of 2-RDM for the reference DSL state. $\rho_{|12\rangle\langle21|} = -|\rho|_{|12\rangle\langle21|}e^{i\delta\phi}$ is complex. The variation of 2-RDM for GSH is shown in Fig. \ref{fig:gsh}. $\Delta|\rho|_{|12\rangle\langle21|}$ increases significantly with increasing $\theta$. So GSH is favored by the twisted permutation term
\begin{eqnarray}
\label{eq:tpm}
H^{GSH}=\sum_{\langle ij\rangle} P_{ij}', \ \ \ \
  P_{ij}'&=&S_\beta^\alpha(i)S_\alpha^\beta(j)e^{i\Theta_{\alpha\beta}},
\end{eqnarray}
with $\Theta_{12}=-\Theta_{21}$  on NN bonds on kagome lattice.  In terms of
spin-1/2 spin operators, the twisted permutation operator contains
Dzyaloshinskii-Moriya (DM) interaction.
GSH has a planar spin order. With large DM interactions, the spin order was
already found in Ref. \onlinecite{Cepas2008} for spin-1/2 systems on the kagome
lattice.

The spin-1 SU(3) DSL  on the kagome lattice has no flux in the mean field
Hamiltonian 
\begin{eqnarray}
  H_{\text{DSL}}=-t\sum_{\langle ij\rangle\alpha}f_{i\alpha}^\dag f_{j\alpha}+\text{h.c.},
\end{eqnarray}
The SU(3) CSL state has the flux $3\theta$ in triangles  $\triangle_{123}$ and
$\triangledown_{1'2'3'}$ and $-6\theta$ in the hexagon; with $\theta=0$ fixed,
we can get the VBS state by varying the ratio of hopping amplitudes $t_2/t_1$.

\begin{figure}[t]
  \begin{center}
    \includegraphics[width=\columnwidth]{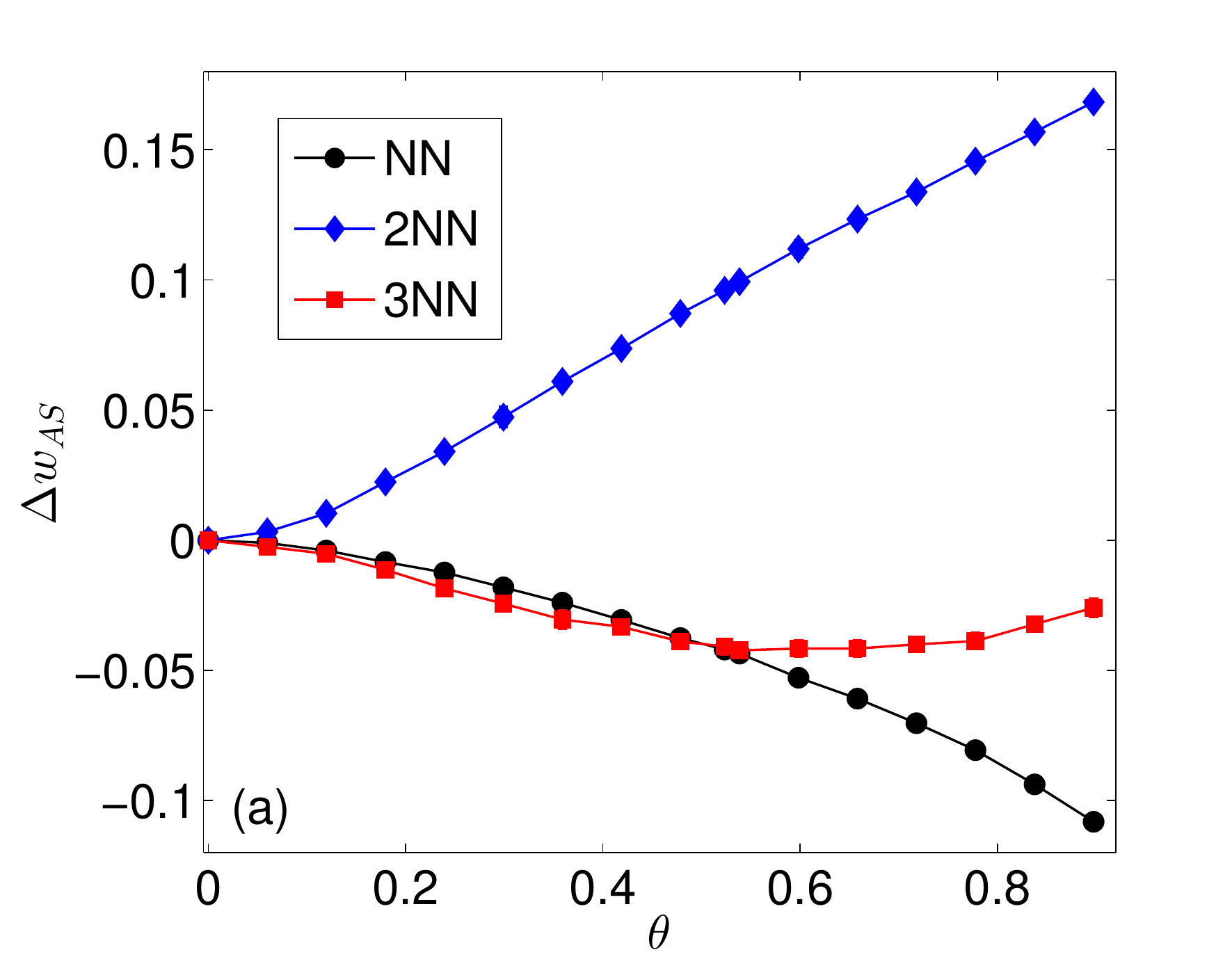}\\
    \includegraphics[width=\columnwidth]{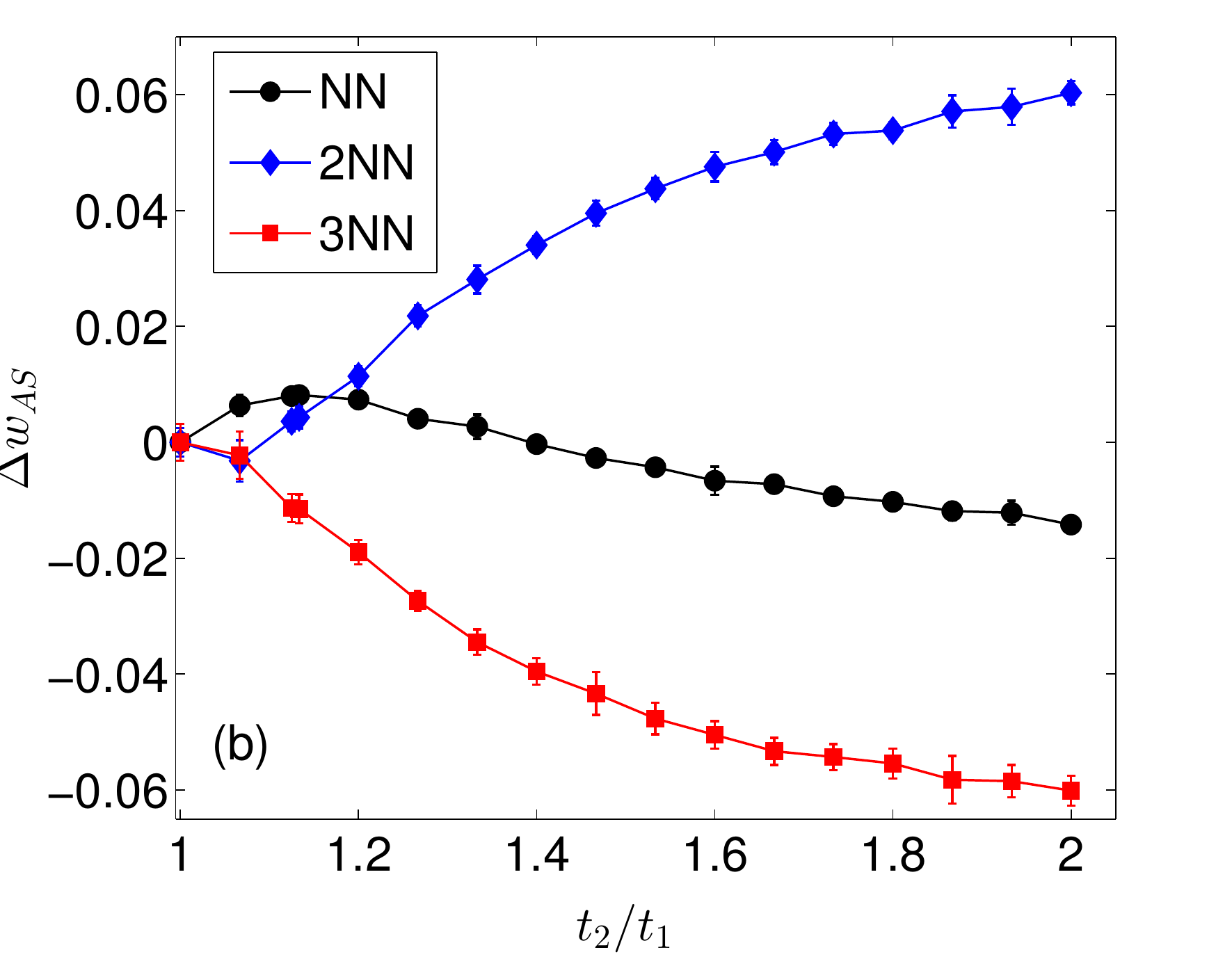}
  \end{center}
  \caption{$\theta$ and $t_2/t_1$-dependent variations of total maximum
anti-symmetric weights $\Delta w_{AS}=w_{AS}-w_{AS}^0$ for SU(3) CSL  (a) and
VBS (b), respectively, on NN, 2NN and 3NN bonds. VBS has the maximum $w_{AS}$
on NN bonds at $t_2/t_1=1.125$ giving the variational energy
$E_{\text{VBS}}=-0.4051(8)J_1$ per bond close to the tensor-network
simulation.\cite{Corboz2012a} We see that only an interaction on 2NN
bonds can favor the SU(3) CSL.}
  \label{fig:su3}
\end{figure}
SU(3) DSL state has 2-RDM of weights $w^0_{\text{AS}}(\text{NN})=0.694(1)$, $w^0_{\text{AS}}(\text{2NN})=0.255(2)$, $w^0_{\text{AS}}(\text{3NN})=0.405(3)$. With increasing parameters ($\theta$ or $t_2/t_1$), the responses of 2-RDM, $\Delta w_{AS}=w_{AS}-w_{AS}^0$ are shown in Fig. \ref{fig:su3}. The nonmonotonicity of $\Delta w_{\text{AS}}(\text{NN})$ for VBS suggests that the NN SU(3) spin model has the stability towards to the VBS state, consistent with the tensor-network numerical calculation.\cite{Corboz2012a} At $t_2/t_1=1.125$, VBS has $w_{\text{AS}}(NN)=0.7025(4)$. The variational energy per bond $E_b^{\text{VBS}}=-0.4051(8)J_1$ very close to ground state energy in tensor-network simulation, $E_g\simeq-0.415J_1$.\cite{Corboz2012a} Both CSL and VBS states are favored by $J_2$ SU(3) term on 2NN bonds. When $J_2/J_1>0.75$, CSL state is the more likely favored state for the SU(3) spin model.

For spin-1 systems, the permutation operator also contains the biquadratic term, $P_{ij}^{S=1}=\mathbf{S}_i\cdot\mathbf{S}_j+(\mathbf{S}_i\cdot\mathbf{S}_j)^2$.
In real materials, the biquadratic coupling is small and the Heisenberg term
dominates\cite{Fazekas1999}, far away from SU(3) limit. However, theoretically
we can still take SU(3) DSL as the reference and turn on the paring of partons
to go back to SO(3) symmetry.\cite{Liu2012} 

\begin{figure}[t]
  \begin{center}
    \includegraphics[width=\columnwidth]{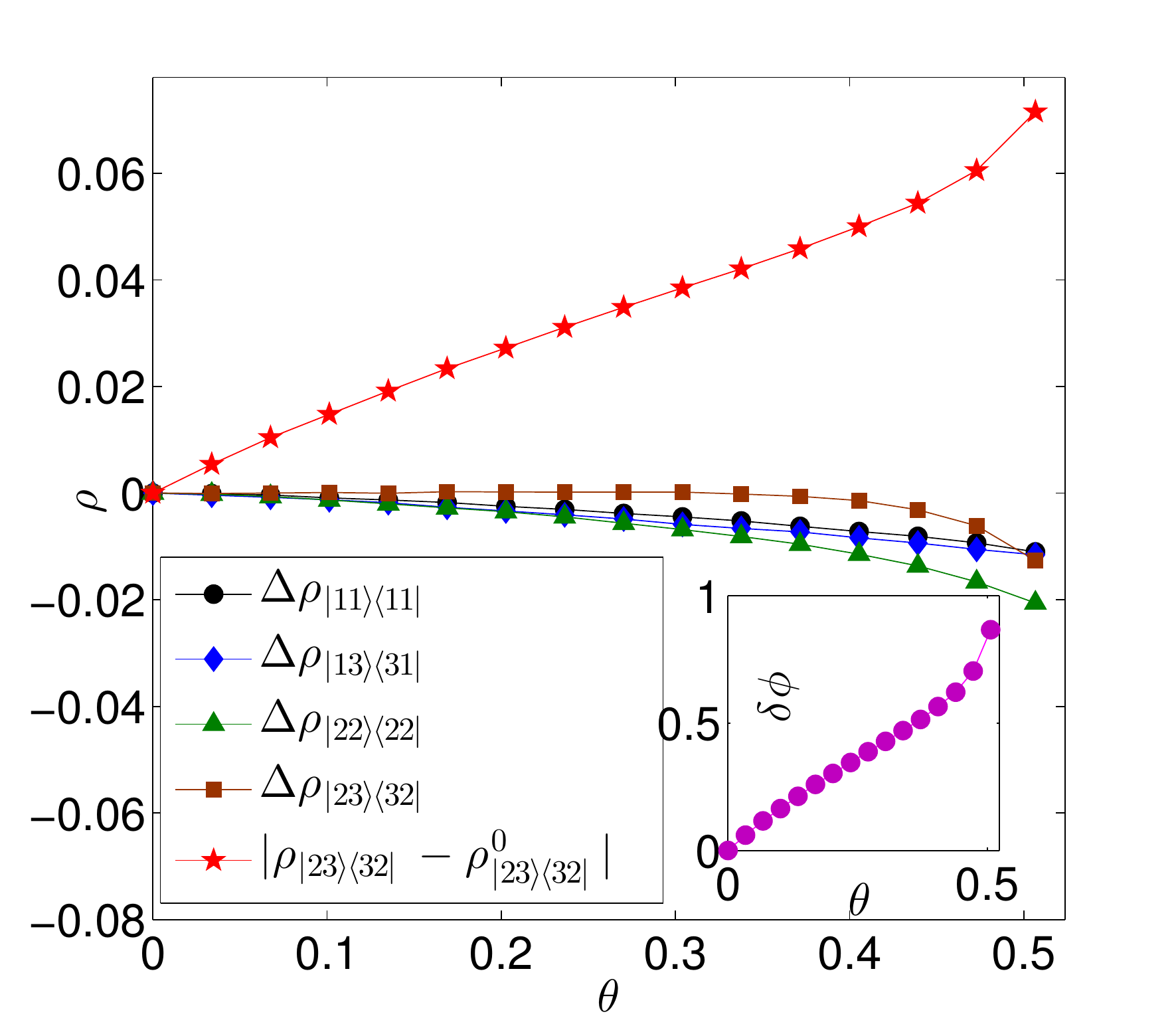}
  \end{center}
  \caption{The $\theta$-dependence of variations of 2-RDM for the SPT state on NN bonds is defined in Eq. (\ref{eq:sptrdm}). The amplitude $|\rho|_{|12\rangle\langle21|}$ varies little and phase $\delta\phi$ increase significantly for the small $\theta$. GSH is potentially stabilized by twisted spin Hamiltonian in Eq. (\ref{eq:sptH}).}
  \label{fig:spt}
\end{figure}
The mean field Hamiltonian of SPT is written as\cite{Lu2012}
\begin{eqnarray}
  H=-t\sum_{\langle ij\rangle}e^{i\theta_\alpha}f_{i\alpha}^\dag f_{j\alpha}+\text{h.c.},
\end{eqnarray}
with $\theta_1=\theta_3=-\theta_2/2=\theta$ ($\theta_1+\theta_2+\theta_3=0$). The SPT state is protected by SU(2)$\times$U(1). The components of 2-RDM, $|11\rangle\langle11|$, $|13\rangle\langle31|$, $|31\rangle\langle13|$ and $|33\rangle\langle33|$, have the SU(2) symmetry. There are four independent weights for 2-RDM, $\rho_{|11\rangle\langle11|}$, $\rho_{|13\rangle\langle31|}$, $\rho_{|22\rangle\langle22|}$ and $\rho_{|23\rangle\langle32|}$. The reference SU(3) DSL has the values
\begin{eqnarray}
 \rho^0_{|11\rangle\langle11|} &=& 0.0509(3), \quad \rho^0_{|13\rangle\langle31|}=-0.0904(5),\nonumber\\
 \rho^0_{|22\rangle\langle22|} &=& 0.0508(2), \quad \rho^0_{|23\rangle\langle32|}=-0.0902(6).\nonumber
\end{eqnarray}
Turning on $\theta$, the variation of 2-RDM is defined by the notations below
\begin{eqnarray}
\label{eq:sptrdm}
 \Delta\rho_{|11\rangle\langle11|} &=& -(\rho_{|11\rangle\langle11|} - \rho^{0}_{|11\rangle\langle11|}),\nonumber\\
 \Delta\rho_{|13\rangle\langle31|} &=& -(\rho_{|13\rangle\langle31|} - \rho^{0}_{|13\rangle\langle31|}),\nonumber\\
 \Delta\rho_{|22\rangle\langle22|} &=& -( \rho_{|22\rangle\langle22|} - \rho^{0}_{|22\rangle\langle22|}),\nonumber\\
\Delta|\rho|_{|23\rangle\langle32|} &=& |\rho|_{|23\rangle\langle32|} - |\rho|^{0}_{|23\rangle\langle32|},\nonumber\\
\delta\phi&=&\arg(-\rho_{|23\rangle\langle32|}), 
\end{eqnarray}
where $\rho_{|23\rangle\langle32|}=-|\rho|_{|23\rangle\langle32|}e^{i\delta\phi}$ is complex.

The $\theta$-dependent 2-RDM of SPT is shown in Fig. \ref{fig:spt}. We see that
the absolute value $|\rho|_{|23\rangle\langle32|}$, and
other $|\rho|_{|ij\rangle\langle ij|}$ vary little ($<0.02$), however, the
phase $\delta\phi \sim 1$ or $|\rho_{|23\rangle\langle32|} -
\rho^{0}_{|23\rangle\langle32|}|\sim 0.1$.  Such a large change in the phase of
$\rho_{|23\rangle\langle32|}$ can be driven by the twisted permutation term
on NN bonds
\begin{eqnarray}
\label{eq:sptH}
H^{SPT}=\sum_{\langle ij\rangle} P_{ij}', \ \ \ \
  P_{ij}'=S_\beta^\alpha(i)S_\alpha^\beta(j)e^{i\Theta_{\alpha\beta}},
\end{eqnarray}
which potentially stabilizes SPT state. Here $\Theta_{12}=\Theta_{32}=\Theta$, $\Theta_{13}=0$ and $\Theta_{\alpha\beta}=-\Theta_{\beta\alpha}$ ($\alpha,\beta=1,2,3$).
We compare the VBS and SPT states for the twisted $J_1$ term and find that SPT will win when $\Theta > 0.6$. More systematic simulations (e.g. DMRG) are needed to confirm the result in further studies.



In conclusion, we start from Gutzwiller-projected parton wave functions to
design supported local spin Hamiltonians. Gutzwiller projective parton
construction is a powerful theoretical approach to construct many different
wave function ansatzs with non-trivial topological properties. Our work
establishes a rough directional guide for further unbiased numerical simulations
for these states.

We thank Zheng-Xin Liu, Peng Ye and Fang-Zhou Liu for helpful discussions.  This research is supported by the BMO Financial
Group and the John Templeton Foundation, and by NSF Grant No.  DMR-1005541 and
NSFC 11274192.  Research at Perimeter Institute is supported by the Government
of Canada through Industry Canada and by the Province of Ontario through the
Ministry of Research.

\bibliography{rdm} 

\newpage 
\onecolumngrid

\section{ Supplemental material: More on SPT}

\begin{figure}[b]
  \begin{center}
    \includegraphics[width=0.75\columnwidth]{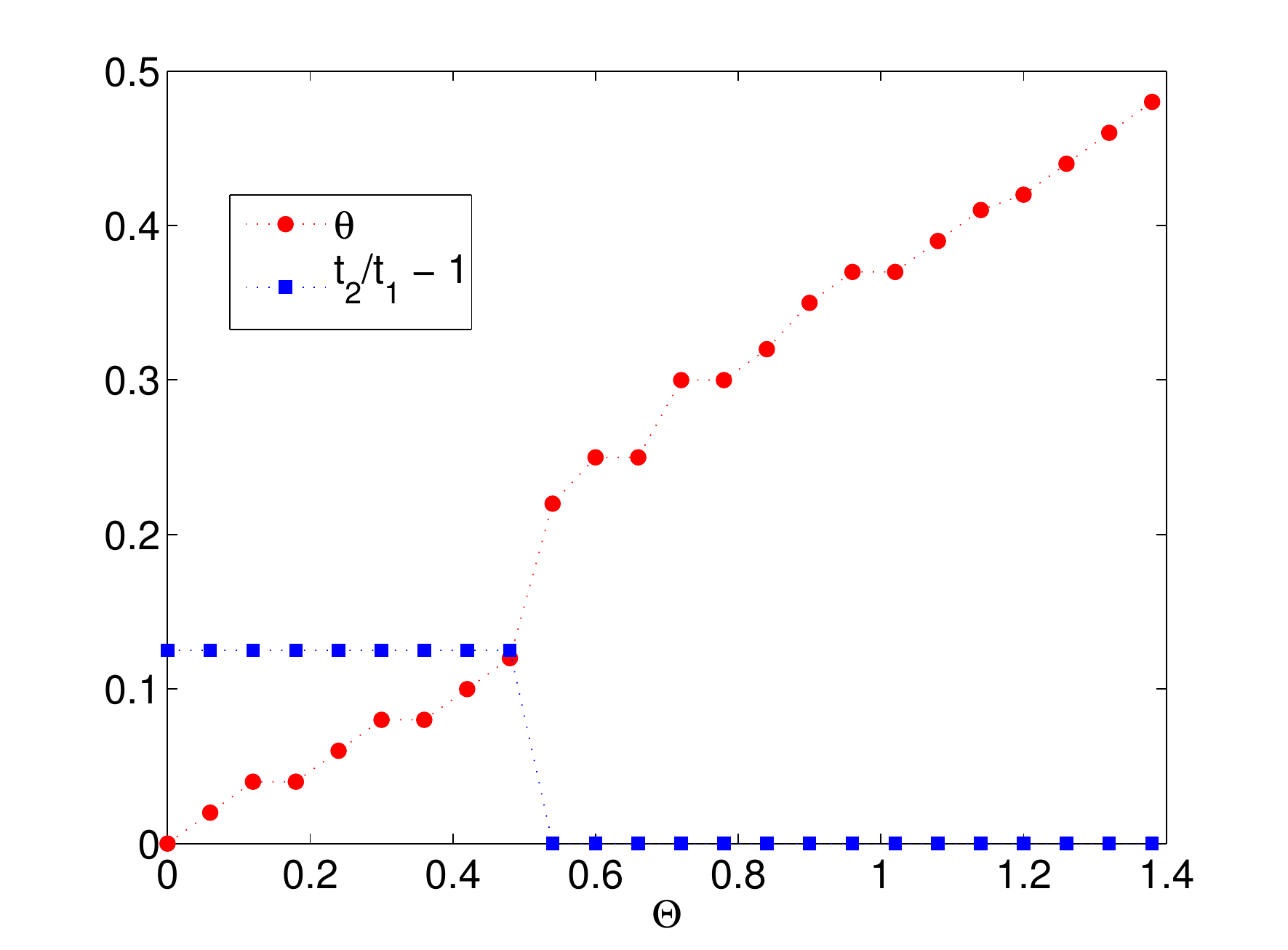}
  \end{center}
  \caption{The $\Theta$ dependence of optimized $t_2/t_1-1$ and $\theta$.}
  \label{fig:su3phase}
\end{figure}
Given a twisted $J_1$ term
\begin{eqnarray}
 P_{ij}^{'S=1}&&=\cos(\Theta)(\mathbf{S}_i\cdot\mathbf{S}_j+(\mathbf{S}_i\cdot\mathbf{S}_j)^2)+\sin(\Theta)\{ [S_i^x(S_j^yS_j^z+S_j^zS_j^y)-(S_i^yS_i^z+S_i^zS_i^y)S_j^x]\nonumber\\
-&&[S_i^y(S_j^xS_j^z+S_j^zS_j^x)-(S_i^xS_i^z+S_i^zS_i^x)S_j^y]\}
+2\sin^2(\Theta/2)[(S_i^xS_i^y+S_i^yS_i^x)(S_j^xS_j^y+S_j^yS_j^x)\nonumber\\
+&&S_i^zS_j^z+((S_i^x)^2-(S_i^y)^2)((S_j^x)^2-(S_j^y)^2)
+\frac{1}{3}(2(S_i^z)^2-(S_i^x)^2-(S_i^y)^2)(2(S_j^z)^2-(S_j^x)^2-(S_j^y)^2)]
\end{eqnarray}
we try to find the ground state in the vicinity of DSL state. We introduce two varational parameters, $t_{2}/t_1$ and $\theta$,  in our mean field Hamiltonian 
\begin{eqnarray}
  H=-\sum_{\langle ij\rangle}t_{ij}e^{i\theta_\alpha}f_{i\alpha}^\dag f_{j\alpha}+\text{h.c.},
\end{eqnarray}
When $\Theta < 0.55$, we always find the finite value of $t_2/t_1-1$ for the ground state. At the same time, the non-zero $\theta$ is also favored. When $\Theta > 0.55$, the uniform hopping amplitude is favored, $t_2/t_1=1$ and there is a jump of the slop for $\theta(\Theta)$.

To confirm the SPT state, we calculate the spin Hall conductance. The SPT state breaks SU(3) symmetry into SU(2)$\otimes$U(1) symmetry. We would like to show that the SPT state here is protected by two U(1) symmetries, $S_z$ and $S_z^2$. The twisted Hamiltonian is indeed $S_z$ and $S_z^2$ invariant. We would calculate the Hall response of SPT respect to $S_z$ and $S_z^2$ U(1) field. We need first to find out the charge assignment for the two different U(1) field. Respect to U(1) response, the mean field Hamiltonian should be written as
\begin{eqnarray}
  H=-t\sum_{\langle ij\rangle}e^{ ia + iq_\alpha {A}}e^{i\theta_\alpha}f_{i\alpha}^\dag f_{j\alpha}+\text{h.c.},
\end{eqnarray}
where $A$ is the external probe U(1) field and $a$ is the internal gauge field to enforce the single-occupation constraint. $\alpha=1,2,3$ flavors have the different Chern number of the mean field filled bands, $C_1=C_3=-C_2=1$. The external $A$ can generate the flavor current
\begin{eqnarray}
  J_\alpha=\frac{C_\alpha}{2\pi} (a+ q_\alpha A),
\end{eqnarray}
To satisfy the single-occupation constraint, we have
\begin{eqnarray}
  \sum_\alpha J_\alpha = 0,
\end{eqnarray}
then we get
\begin{eqnarray}
  a=-\frac{\sum_{\alpha}C_\alpha q_\alpha}{\sum_\alpha C_\alpha}A
\end{eqnarray}
Therefore, the response mean field Hamiltonian is given as
\begin{eqnarray}
  H=-t\sum_{\langle ij\rangle}e^{ iQ_\alpha {A}}e^{i\theta_\alpha}f_{i\alpha}^\dag f_{j\alpha}+\text{h.c.}
\end{eqnarray}
with the effective charge respect to $A$, $Q_\alpha=q_\alpha-\frac{\sum_{\alpha}C_\alpha q_\alpha}{\sum_\alpha C_\alpha}$. For $S_z$, we get $\mathbf{q}=(1, 0, -1)$ and $\mathbf{Q}=(1, 0, -1)$. For $S_z^2$, $\mathbf{q}=(1,0,1)$ and $\mathbf{Q}=(-1,-2,-1)$. All the charges are integers. Therefore there is no  ground state degeneracy. To see this clearly, we can also calculate $S_z$ and $S_z^2$ charges for chiral spin liquid state. For CSL, $\mathbf{Q}=(1,0,-1)$ for $S_z$ and $\mathbf{Q}=(\frac{1}{3},-\frac{2}{3},\frac{1}{3})$ for $S_z^2$. From the charge fractionalization, we know CSL has the three-fold ground state degeneracy.

\begin{figure}[t]
  \begin{center}
    \includegraphics[width=0.75\columnwidth]{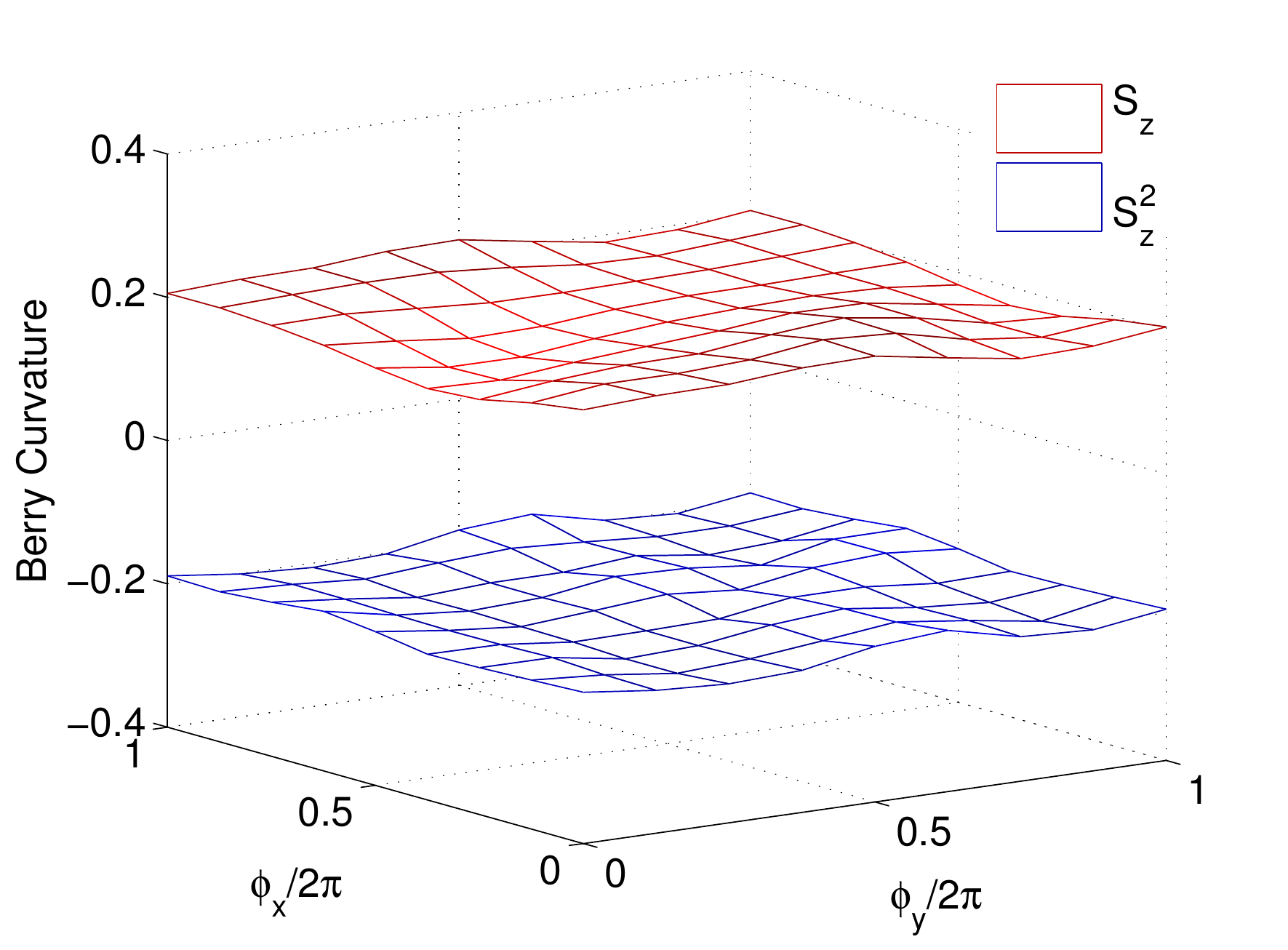}
  \end{center}
  \caption{The spin Berry curvature for $S_z$ and $S_z^2$ for SPT state.}
  \label{fig:su3berry}
\end{figure}
To calculate the spin Hall conductance, we impose the general twisted boundary phase on the system 
\begin{eqnarray}
  f_{i+L_x\alpha}=f_{i\alpha}e^{iQ_\alpha \phi_x}, \quad f_{i+L_y\alpha}=f_{i\alpha}e^{iQ_\alpha \phi_y}.
\end{eqnarray}
From the response mean field Hamiltonian, we can obtain the response Gutzwiller-projected wave function. The spin Hall conductance is given as
\begin{eqnarray}
  \sigma_s=\frac{1}{2\pi}\int_{-\pi}^{\pi}d\phi_x\int_{-\pi}^{\pi}d\phi_y\Omega(\phi_x,\phi_y)
\end{eqnarray}
with the Berry curvature
\begin{eqnarray}
  \Omega(\phi_x,\phi_y)=\nabla\times\langle\Psi_G|i\nabla_\phi|\Psi_G\rangle
\end{eqnarray}
 Numerically, we divide the unit cell of the boundary phases into 8-by-8 mesh points. The spin Hall conductance is given as
\begin{eqnarray}
  \sigma_s=\frac{1}{2\pi}\sum_{j}\Omega_j
\end{eqnarray}
with Berry curvature
\begin{eqnarray}
  \Omega_j=\arg\prod_i\langle \Psi_G^{j_{i+1}}|\Psi_G^{j_i}\rangle,
\end{eqnarray}
where $i =1-4$ (with $j_5\equiv j_1$) denote four mesh points at the $j$-th plaquette of the $k$ mesh patches.
The Gutzwiller projected wavefunctions can be written as
\begin{eqnarray}
  |\Psi\rangle = \sum_{R_i}\psi(R_i)|R_i\rangle
\end{eqnarray}
with
\begin{eqnarray}
  \psi(R_i) = \det(\varphi_i(\mathbf{r}_i))
\end{eqnarray}
On the $k$ mesh, we need calculate $2N_k$ overlap of the wavefunctions
\begin{eqnarray}
 \frac{\langle \Psi_j|\Psi_{j'}\rangle}{\langle\Psi_j|\Psi_j\rangle} &=& \frac{\sum_{R_i}\psi_j^*(R_i)\psi_{j'}(R_i)}{\sum_{R_i}|\psi_j(R_i)|^2}\nonumber\\
&=&\frac{\sum_{R_i}|\psi_j(R_i)|^2\frac{\psi_{j'}(R_i)}{\psi_j(R_i)}}{\sum_{R_i}|\psi_j(R_i)|^2}
\end{eqnarray}
where the summation can be done with the standard VMC technique with the weight $|\psi_j(R_i)|^2$. To calculate the berry curvature, we need calculate $2N_k$ overlap of the wavefunctions with different boundary condition $\mathbf{k}$. Thus we will take the weight
\begin{eqnarray}
  \rho(R_i) = \sum_{k}|\psi_k(R_i)|^2
\end{eqnarray}
and the overlap is calculated as
\begin{eqnarray}
  \frac{\langle\Psi_j|\Psi_{j'}\rangle}{\sum_{R_i}\rho(R_i)} &=& \frac{\sum_{R_i}\psi^*_j(R_i)\psi_{j'}(R_i)}{\sum_{R_i}\rho(R_i)}\nonumber\\
  &=&\frac{\sum_{R_i}\rho(R_i)\frac{\psi_j^*(R_i)\psi_{j'}(R_i)}{\rho(R_i)}}{\sum_{R_i}\rho(R_i)}
\end{eqnarray}

The spin Berry curvature respect to $S_z$ and $S_z^2$ for SPT are shown in Fig. \ref{fig:su3berry}. The spin Hall conductance is $\sigma_s(S_z)=2$ and $\sigma_s(S_z^2)=-2$. It is the same as the results from $K$-matrix methods.

\end{document}